\documentclass[twocolumn,showpacs,preprintnumbers,amsmath,amssymb,eqsecnum]
{revtex4}
\usepackage{bm}% bold math
\begin{document}

%\preprint{}

\title{On the Electric-Dipole Representation in Atomic Physics}
% Force line breaks with \\

\author{Francesco Miglietta}
\affiliation{Dipartimento di Fisica dell' Universit\`a
 di Pavia - via Bassi 6, I-27100 Pavia - Italy }
% \\ I.N.F.N. - Sezione di Pavia 

\email{francesco.miglietta@pv.infn.it}
\date{October 3, 2012}

\begin{abstract}
The unitary transformation that leads from the minimal-coupling description to
the electric-dipole one is analysed in detail. The momentum cut-off function  $
\hat{f}(k) $, which is understood in the definition of such a transformation, is
obtained explicitly by use of the variational method. We give an expression  for
$ \hat{f}(k) $  in terms of the electron-density and of the mean-squared value
of the electric-dipole moment of the atom in its ground state. A relevant consequence concerns the van der Waals interaction, whose long-distance 
behaviour turns out to be proportional to $ R^{-8} $.
\end{abstract}

\maketitle

\section{Introduction}

In non-relativistic quantum mechanics the standard description for an atom
interacting with the radiation field is provided by the minimal-coupling (MC)
Hamiltonian. As it is well known, the MC Hamiltonian can be obtained in a 
natural way as a non-relativistic limit for the Dirac Hamiltonian (see  {\it
e.g.} Ref. \cite{1} ). Nevertheless it is known also that, in the
long-wavelength limit, the gauge coupling of the MC description is replaced
usually  by the electric-dipole (ED) coupling $ -\bm{d} \cdot \bm{E} $.  \par

The transition from one description to the other one is analysed, in several
books, in terms of classical electromagnetic field (see {\it e.g.} Ref. \cite{2}
\cite{3} ). Much more elegant is a method in which, from the beginning, the
electromagnetic field is introduced as a quantum field (see {\it e.g.} Ref. 
\cite{4}). Such a procedure is based on the introduction of a unitary
transformation, which leads from the MC representation to the ED one. Some
interesting features concerning such a transformation will be analysed in this
paper.  \par

Let us introduce the problem briefly. The total Hamiltonian $ H_T $, which
describes an atom interacting with the electromagnetic field, consists of
%1.1
\begin{equation}
H_T = H_F + H'_{M} .
\end{equation}
\par
In eq.(1.1) the free-field Hamiltonian $ H_F $ is given, in Gaussian units and 
in the Coulomb gauge, by
%1.2
\begin{eqnarray}
 H_F && = \frac{1}{8 \pi} \int d \bm{r} :( E^2_{\bot} + B^2 ): \\ \nonumber
&& = \sum _{\bm{k} \alpha } \hbar \omega_k 
{\alpha }^+_{\bm{k} \alpha } {\alpha }_{\bm{k} \alpha}
\end{eqnarray}
(semicolons mean normal ordering). The Hamiltonian $ H'_M $, which describes 
both the atom and its MC interaction with the radiation field, has the 
well-known form
%1.3 
\begin{equation}
H'_M = \frac{1}{2m} : \vert -i \hbar \bm{\nabla } + \frac{e}{c} \bm{A} (\bm{r} )
{\vert }^2 : + V(\bm{r} - \bm{R} ).
\end{equation}

For the sake of simplicity we assume a single electron, we neglect the spin and
we assume that the centre-of-mass of the atom be fixed at the position $
\bm{R}$.  \par In eq.(1.3) the normal ordering, which concerns the $ A^2 $ term
only, has the effect of fixing a reference-level for the energy, according to
the following relation
%1.4
\begin{equation}
\langle  g |\langle  0| H'_M |0 \rangle |g \rangle = \langle  g|H_{M0} |g
\rangle = E_0.
\end{equation}
In eq.(1.4) $ |0 \rangle $ is the vacuum state for the field,
%1.5
\begin{equation}
H_{M0} = -\frac{{\hbar }^2 }{2m} {\nabla }^2 + V(\bm{r} - \bm{R})
\end{equation}
is the Hamiltonian for the unperturbed atom and $ E_0 $ is the energy-eigenvalue
corresponding to the ground-state $ |g \rangle $.  \par

The transition to the ED representation is accomplished through the following
unitary transformation \cite{4}
%1.6
\begin{equation}
U = \exp [ - \frac{i}{ \hbar c} \bm{d} \cdot \bm{A} (\bm{R} ) ] ,
\end{equation}
where a high-momentum cut-off is understood implicitly.    \par

In this paper we will proceed in the following way. First of all we will
introduce a cut-off function $ \hat{f}(k) $, for the photon momentum $ k $,
explicitly in the definition of the unitary transformation $ U $ of eq.(1.6).
This will be done in eq.s (2.1) and (2.3). Subsequently the function  $
\hat{f}(k) $will be determined by the variational method. As a result we will
obtain for $ \hat{f}(k) $ an expression involving the electron-density $
\hat{n}_e (k) $ and the mean-squared value $ \langle  d^2 \rangle $ of the ED
moment of the atom (see eq.(3.8) in the sequel). In the ED representation the
interaction-Hamiltonian will contain the cut-off function $ \hat{f}(k) $
explicitly and it will differ significantly from the simple expression
$ -\bm{d} \cdot \bm{E}_{\bot } (\bm{R} )$, which is used currently.   \par

The case of $N$ atoms is treated in Sec.IV . The unitary transformation is
redefined suitably in eq.(4.1). In this case the transformation gives rise to  a
dipole-dipole interaction among different atoms, to be ascribed to the
transverse field. Such an interaction overlaps with the dipole-dipole 
interaction due to the Coulomb field. As a result, at large distance an exact
cancellation of the contributions proportional to $ R^{-3} $ occurs. So the
leading contribution to the large-distance dipole-dipole interaction turns out 
to be proportional to $ R^{-4} $. This would imply for the van der Waals
potential a leading contribution proportional to $ R^{-8} $ and not to $ R^{-6}
$. This is shown in Sec.V. Actually the experimental status does not look
conclusive.

\section{The Electric-Dipole representation}

The transition to the ED representation is achieved through the following 
unitary transformation (Gaussian units are used)
%2.1
\begin{equation}
U = \exp [- \frac{i}{\hbar c} \bm{d} \cdot {\bm{A} }^f (\bm{R} ) ]  ,
\end{equation} 
where
%2.2
\begin{equation}
\bm{d} = -e (\bm{r} - \bm{R} )
\end{equation}
is the ED operator. In eq.(2.1) a momentum cut-off has been introduced, by
defining 
%2.3
\begin{eqnarray}
{\bm{A} }^f (\bm{r} ) &=& c \sqrt{\frac{2 \pi \hbar }{V} } \sum _{\bm {k}
 \alpha} \frac{\hat{f}(k) }{ \sqrt{{\omega}_k }} [{\bm{e} }_{\bm{k} \alpha}
 a_{\bm{k} \alpha} e^{i \bm{k} \cdot \bm{r} } 
  \\ \nonumber  
&& + {\bm{e} }^*_{\bm{k} \alpha} a^+_{\bm{k} \alpha}
 e^{-i \bm{k} \cdot \bm{r} } ] .
\end{eqnarray}
In a similar way we define
%2.4
\begin{eqnarray}
{\bm{E} }^f_{\bot } (\bm{r} ) &=& i \sqrt{\frac{2 \pi \hbar }{V} } 
 \sum _{\bm{k} \alpha} \hat{f} (k) \sqrt{{\omega}_k } [\bm{e}_{\bm{k}   
 \alpha} a_{\bm{k} \alpha} e^{i \bm{k} \cdot \bm{r} }  \\ \nonumber
&& - {\bm{e} }^*_{\bm{k} \alpha}a^+_{\bm{k} \alpha}
 e^{-i \bm{k} \cdot \bm{r} } ] .
\end{eqnarray}
The {\it true} field operators $\bm{A} $ and $\bm{E}_{\bot} $  correspond to $
\hat{f} = 1 $.  \par The real isotropic function $\hat{f}(k) $ carries out the
high-momentum cut-off. It will be determined in Sec.III by the variational
method.  \par The presence of the cut-off function $\hat{f}(k) $ affects the
commutation relations among field operators, namely
%2.5
\begin{eqnarray}
&& [A^f_i (\bm{R} ) , E_{\bot j} (\bm{r} ) ]   \\ \nonumber
&& = i 4 \pi \hbar c V^{-1} \sum_{\bm{k} } e^{i \bm{k} \cdot (\bm{r} - \bm{R} )}
\hat{f} (k) ( {\delta }_{ij} - \frac{k_i k_j }{k^2 } ).
\end{eqnarray}  \par
Owing to this result, the transverse electric field operator transforms as 
%2.6
\begin{eqnarray}
&& U  E_{\bot j} (\bm{r} ) U^{-1}  \\ \nonumber
&& = E_{\bot j} (\bm{r} )  - \frac{i}{\hbar c} \sum_i d_i \:
  [ A^f_i (\bm{R} ) , E_{\bot j} (\bm{r} ) ]  \\ \nonumber
&& = E_{\bot j} (\bm{r} ) - \frac{4 \pi }{V} \sum_{\bm{k} i} d_i \:
e^{i \bm{k} \cdot ( \bm{r} - \bm{R} )} \hat{f}(k) 
( {\delta }_{ij} - \frac{k_i k_j }{k^2 } )  .
\end{eqnarray}  \par
The vector-potential $\bm{A} $, as well as the magnetic-induction $\bm{B} $, is
not modified by the transformation. These results entail that, in the ED
representation, the free-field Hamiltonian $ H_F $ reads
%2.7
\begin{eqnarray}
&& U H_F U^{-1}  = H_F   \\ \nonumber
&&  - \sum_i d_i \, V^{-1} \sum_{\bm{k} i} \int d \bm{r} \,
E_{\bot j} (\bm{r} )
e^{i \bm{k} \cdot ( \bm{r} - \bm{R} )} \hat{f}(k) 
( {\delta }_{ij} - \frac{k_i k_j }{k^2 } )  \\ \nonumber
&& + 2 \pi \sum_{ij} d_i d_j \, V^{-1} \sum_{\bm{k} }
{\hat{f} }^2 (k) ( {\delta }_{ij} - \frac{k_i k_j }{k^2 } )  .
\end{eqnarray} 
By calculating the integral in the second term we obtain
%2.8
\begin{eqnarray}
&& V^{-1} \, \sum_{\bm{k} j}\int d \bm{r} \, E_{\bot j} (\bm{r} )
e^{i \bm{k} \cdot ( \bm{r} - \bm{R} )} \hat{f}(k) 
( {\delta }_{ij} - \frac{k_i k_j }{k^2 } )   \\ \nonumber
&& = i \sqrt{\frac{2 \pi \hbar }{V} } \sum_{\bm{k} \alpha} \hat{f}(k) 
\sqrt{{\omega }_k } [\bm{e}_{\bm{k} \alpha} a_{\bm{k} \alpha} 
e^{i \bm{k} \cdot \bm{R} }  - {\bm{e} }^*_{\bm{k} \alpha}
a^+_{\bm{k} \alpha} e^{-i \bm{k} \cdot \bm{R} } ]  \\ \nonumber
&& = E^f_{\bot i} (\bm{R} ).
\end{eqnarray}
The integral in the third term gives 
%2.9
\begin{eqnarray}
&& 2 \pi V^{-1} \sum_{\bm{k} }
 {\hat{f} }^2 (k) ( {\delta }_{ij} - \frac{k_i k_j }{k^2 } )  \\ \nonumber
&& \simeq \frac{1}{4 {\pi }^2 } \int d \bm{k} \, {\hat{f} }^2 (k) 
 [\frac{2}{3} {\delta }_{ij} - (\frac{k_i k_j }{k^2 } - \frac{1}{3} 
 {\delta }_{ij} ) ]  \\ \nonumber
&& = \frac{2}{3\pi } {\delta }_{ij} \int_0^{\infty } dk \, k^2 
 {\hat{f} }^2 (k).
\end{eqnarray}
In performing the calculation, the scalar contribution ($L=0$) has been 
separated from the null-trace one ($L=2$).  \par

In this way we obtain
%2.10
\begin{eqnarray}
&& U H_F U^{-1}  \\ \nonumber
&& = H_F - \bm{d} \cdot {\bm{E} }^f_{\bot } (\bm{R} ) + \frac{2}{3 \pi }
 | \bm{d} |^2 \int_0^{\infty } dk \, k^2 {\hat{f} }^2 (k).
\end{eqnarray}
Finally, putting all these results together, we obtain
%2.11
\begin{eqnarray}
&& U H_T U^{-1}  \\ \nonumber
&& = H_F + \frac{1}{2m} : \vert -i \hbar \bm{\nabla } + \frac{e}{c} 
 [ \bm{A} (\bm{r} )  - {\bm{A} }^f (\bm{R} )] {\vert }^2 : \\ \nonumber
&& + V(\bm{r} -\bm{R} )  
 - \bm{d} \cdot {\bm{E} }^f_{\bot } (\bm{R} ) + \frac{2}{3 \pi }
 | \bm{d} |^2 \int_0^{\infty } dk \, k^2 {\hat{f} }^2 (k)  \\ \nonumber
&& + \frac{e^2}{ 2m c^2 } \langle  0| \vert {\bm{A} }^f (\bm{R} ) {\vert }^2 
|0 \rangle   \\ \nonumber
&& - \frac{e^2 }{m c^2 } \langle  0|\bm{A} (\bm{r} ) \cdot 
 {\bm{A} }^f (\bm{R} )|0 \rangle  .
\end{eqnarray}

We have to explicitate the last two terms. From the first one we obtain
%2.12
\begin{eqnarray}
&& \frac{e^2}{ 2m c^2 } \langle  0| \vert {\bm{A} }^f (\bm{R} ){\vert }^2 
|0 \rangle   \\ \nonumber
&& = 2 \pi e^2 {\lambda }_c V^{-1} \sum_{\bm{k}} k^{-1} 
 {\hat{f} }^2 (k)     \simeq \frac{e^2 {\lambda }_c }{\pi }
 \int_0^{\infty }  dk \, k {\hat{f} }^2 (k) ,
\end{eqnarray}
where $ {\lambda }_c = \hbar / mc $ is the Compton wave-length for the electron.
From the second term we obtain
%2.13
\begin{eqnarray}
&& - \frac{e^2 }{m c^2 } \langle  0|\bm{A} (\bm{r} ) \cdot 
 {\bm{A} }^f (\bm{R} )|0 \rangle   \\ \nonumber
&& = - 4 \pi e^2 {\lambda }_c V^{-1} \sum_{\bm{k}} k^{-1} {\hat{f} } (k) 
 e^{i \bm{k} \cdot (\bm{r} - \bm{R} )}   \\ \nonumber
&& \simeq -\frac{ e^2 {\lambda }_c }{2 {\pi }^2 } \int d \bm{k} \, 
 k^{-1} \hat{f} (k) e^{i \bm{k} \cdot (\bm{r} - \bm{R} )}  .
\end{eqnarray}

In this way from eq.(2.11) we obtain
%2.14
\begin{eqnarray}
&& U H_T U^{-1}  \\ \nonumber
&& = H_F + \frac{1}{2m} : \vert -i \hbar \bm{\nabla } + \frac{e}{c} 
 [ \bm{A} (\bm{r} )  - {\bm{A} }^f (\bm{R} )] {\vert }^2 : \\ \nonumber
&& + V(\bm{r} -\bm{R} )  
 - \bm{d} \cdot {\bm{E} }^f_{\bot } (\bm{R} ) + \frac{2}{3 \pi }
 | \bm{d} |^2 \int_0^{\infty } dk \, k^2 {\hat{f} }^2 (k) \\ \nonumber
&& +  \frac{e^2 {\lambda }_c }{\pi } \int_0^{\infty }  dk \, k 
 {\hat{f} }^2 (k) -\frac{ e^2 {\lambda }_c }{2 {\pi }^2 } \int d \bm{k} \, 
 k^{-1} \hat{f} (k) e^{i \bm{k} \cdot (\bm{r} - \bm{R} )}  .
\end{eqnarray}

Currently eq.(2.14) is written with $ \hat{f} = 1 $, and the last two terms (the
first of which would diverge for $ \hat{f} = 1 $, while the second  one would
give the result $ -2 e^2 {\lambda }_c {\pi }^{-1} \\ \vert \bm{r}  - \bm{R}
{\vert }^{-2} $) are neglected. In the $ \vert \bm{d} {\vert }^2 $  term, which
diverges also for $ \hat{f} = 1 $, a momentum cut-off is understood.
Nevertheless eq.(2.14) differs from the expression
%2.15
\begin{equation}
U H_T U^{-1} = H_F + H_{M0} - \bm{d} \cdot {\bm{E} }_{\bot } (\bm{R} ) ,
\end{equation}
which is assumed usually.  \par

In the sequel we will refer to eq.(2.14), where the function $ \hat{f} $ will be
determined in Sec.III by the variational method. \par

Let us proceed to a preliminary analysis of eq.(2.14). By averaging both sides 
of eq.(2.14) over the vacuum state of the field we obtain
%2.16
\begin{eqnarray}
&& {H'}_{M0} \equiv \langle  0| U H_T U^{-1} |0 \rangle      \\ \nonumber
&& = H_{M0} +\frac{e^2 {\lambda }_c }{\pi }\int_0^{\infty }  
 dk \, k {\hat{f} }^2 (k)     \\ \nonumber
&& +  \frac{2}{3 \pi }
 |\bm{d} |^2 \int_0^{\infty } dk \, k^2 {\hat{f} }^2 (k) \\ \nonumber
&& -\frac{ e^2 {\lambda }_c }{2 {\pi }^2 } \int d \bm{k} \, 
 k^{-1} \hat{f} (k) e^{i \bm{k} \cdot (\bm{r} - \bm{R} )} .
\end{eqnarray}
This equation can be generalised easily to the case of a many-electron atom. 
\par

The hydrogen-atom case is simple. In fact in such a case eq.(2.16) reads
%2.17
\begin{equation}
{H'}_{M0} = H_{M0} + {\cal{V} }' (\vert \bm{r} - \bm{R} \vert ),
\end{equation}
where the additional potential $ {\cal{V} }' $ has the  form
%2.18
\begin{equation}
{\cal{V} }' = C_0 + \frac{1}{2} m {\Omega }^2_0 \vert \bm{r} - \bm{R} {\vert }^2
+  {\cal{V} }" (\vert \bm{r} - \bm{R} \vert ) .
\end{equation}
In eq.(2.18) $ C_0 $ is the following constant
%2.19
\begin{equation}
C_0 = {\pi }^{-1} e^2 {\lambda }_c \int_0^{\infty } dk \, k {\hat{f} }^2 ( k ) ,
\end{equation}
while, for the harmonic-oscillator potential in the second term, one has
%2.20
\begin{equation}
\frac{1}{2} m {\Omega }^2_0 = \frac{2 e^2 }{3 \pi } \int_0^{\infty } dk \, k^2
{\hat{f} }^2 ( k )  .
\end{equation}
Finally the last term consists of the following additional potential
%2.21
\begin{eqnarray}
&& {\cal{V} }" (r) = - \frac{ e^2 {\lambda }_c }{2 \pi } \int d \bm{k} \,
 k^{-1} \hat{f} (k) e^{i \bm{k} \cdot \bm{r} }  \\ \nonumber
&& = - \frac{2 e^2 {\lambda }_c }{\pi r} \int_0^{\infty } dk \, \hat{f} (k)
 \sin kr  \sim \frac{2 e^2 {\lambda }_c }{\pi r^2 }   ,
\end{eqnarray}
where the last side represents the asymptotic expression for large $r$. In the
calculation of the last expression the result $\hat{f} (0) =1 $ has been used.
This result will be obtained in the sequel (see eq.(3.9)). We observe that  $
{\cal{V} }" (r) $ does not vanish for $ r = 0 $. In fact it is
%2.22
\begin{equation}
{\cal{V} }" (0) = - 2 {\pi }^{-1} e^2 {\lambda }_c \int_0^{\infty } dk \, k
 \hat{f} (k)  .
\end{equation} 

In the case of a $Z$-electron atom, eq.(2.2) is replaced by
%2.23
\begin{equation}
\bm{d} = - e \sum^Z_{A=1} ( {\bm{r}}_A -\bm{R} )
\end{equation}
and eq.(2.14) reads
%2.24
\begin{eqnarray}
&& U H_T U^{-1}  \\ \nonumber
&& = H_F + \frac{1}{2m} \sum_A : \vert -i \hbar {\bm{\nabla }}_A 
 + \frac{e}{c} [ \bm{A} ({\bm{r}}_A )
 - {\bm{A} }^f (\bm{R} )] {\vert }^2 : \\ \nonumber
&& + V^{(1)} ({\bm{r}}_A -\bm{R}) + \sum_{A < B} V^{(2)}
 ({\bm{r}}_A - {\bm{r}}_B )   \\ \nonumber
&& - \bm{d} \cdot {\bm{E} }^f_{\bot } (\bm{R} ) + \frac{2}{3 \pi }
 | \bm{d} |^2 \int_0^{\infty } dk \, k^2 {\hat{f} }^2 (k) \\ \nonumber
&& +  {\pi }^{-1} e^2 {\lambda }_c Z  \int_0^{\infty }  dk \, k 
 {\hat{f} }^2 (k)  \\ \nonumber
&& -\frac{ e^2 {\lambda }_c }{2 {\pi }^2 } \sum_A \int d \bm{k} \, 
 k^{-1} \hat{f} (k) e^{i \bm{k} \cdot ({\bm{r}}_A - \bm{R} )}  .
\end{eqnarray}

\section{Variational calculation}

We will determine the function $ \hat{f} $ by the variational method, assuming a
{\it trial} state of the form
%3.1
\begin{equation}
| {\psi }_T \rangle = U |0 \rangle |g \rangle  ,
\end{equation}
where $ |g \rangle $ is the ground-state for the unperturbed atom. The
corresponding wave-function will be indicated as
%3.2
\begin{equation}
{\phi }_0 (\bm{r} -\bm{R}) = \langle  \bm{r} |g \rangle   .
\end{equation}
We observe that the {\it trial} state of eq.(3.1) is normalized. For the sake of
simplicity we assume that the wave-function $ {\phi }_0 $ is isotropic ($ l=0 $)
and we neglect the spin.  \par The energy of the system, averaged over the state
(3.1),
%3.3
\begin{equation}
{\cal{E}}_0 [\hat{f}] = \langle  {\psi }_T |H_T |{\psi }_T  \rangle  
 = \langle  g|\langle  0|U^{-1} H_T U |0 \rangle|g \rangle  ,
\end{equation}
depends on $ \hat{f} $ functionally. Using eq.(2.16) we can write
%3.4
\begin{eqnarray}
&& {\cal{E}}_0 [\hat{f}] = E_0 + {\pi }^{-1} e^2 \{ {\lambda }_c 
 \int_0^{\infty }  dk \, k {\hat{f} }^2 (k)  \\ \nonumber
&& + \frac{2}{3} \frac{\langle  d^2 \rangle}{e^2 } \int_0^{\infty }  dk \, k^2 
 {\hat{f} }^2  (k) - 2 {\lambda }_c \int_0^{\infty }  dk \, k
 {\hat{n} }_e (k) \hat{f} (k) \} .
\end{eqnarray}
In eq.(3.3) $ \langle   d^2 \rangle $ is given by
%3.5
\begin{eqnarray}
&& \langle   d^2 \rangle = e^2 \int d \bm{r} \, r^2 
 \vert {\phi }_0 (r) {\vert }^2    \\ \nonumber
&& = 4 \pi e^2 \int_0^{\infty } dr \, r^4 
 \vert {\phi }_0 (r) {\vert }^2  ,
\end{eqnarray}
while
%3.6
\begin{eqnarray}
&& {\hat{n} }_e (k) = \int d \bm{r} \, e^{i \bm{k} \cdot \bm{r} } 
 \vert {\phi }_0 (r) {\vert }^2    \\ \nonumber
&& = 4 \pi k^{-1} \int_0^{\infty } dr \, r \vert {\phi }_0 (r) {\vert }^2 
 \sin kr   
 \simeq 1 - \frac{ \langle  d^2 \rangle }{6 e^2 } k^2 + ...
\end{eqnarray}
consists of the Fouri\'er-transform of the electron-density $ n_e (r) = \vert
{\phi }_0 (r) {\vert }^2 $. The last side represents the asymptotic behavior
for  $ k \rightarrow 0 $.    \par

Eq.(3.4) confirms the need for a high-momentum cut-off. In fact it can be seen
immediately that for $ \hat{f} = 1 $ the energy $ {\cal{E} }_0 $ diverges.  \par

The function $ \hat{f} (k) $ can be obtained by minimising  the energy $
{\cal{E} }_0 $ of eq.(3.4). We observe that a vanishing $ \hat{f} $ would return
the result $ {\cal{E} }_0 = E_0 $, in agreement with eq.(1.4). Therefore it can
be expected that any non-trivial solution $ \hat{f} $ to the variational problem
would correspond to an energy $ {\cal{E} }_0 <  E_0 $.   \par

By minimising the energy $ {\cal{E} }_0 [ \hat{f} ]$ of eq.(3.4) with respect
to  $ \hat{f} (k) $ we obtain the equation 
%3.7
\begin{eqnarray}
&& 0 = \delta {\cal{E} }_0 [ \hat{f} ] / \delta \hat{f} (k)  \\ \nonumber
&& = \frac{2}{\pi } k e^2 [ ({\lambda }_c + \frac{2 \langle  d^2 \rangle}
{3 e^2 } k ) \hat{f} (k) - {\lambda }_c {\hat{n} }_e (k) ]  .
\end{eqnarray}
The solution to this equation is given by
%3.8
\begin{equation}
\hat{f} (k) = {\hat{n} }_e (k)( 1
 + \frac{2 \langle  d^2 \rangle }{3 e^2 {\lambda }_c } k )^{-1}   .
\end{equation} 

It can be seen immediately that $ \hat{f} (0) = 1 $, as it could be expected.
Furthermore the following asymptotic behaviour for small $ k $  
%3.9
\begin{equation}
\hat{f} (k) \sim 1 - \frac{2 \langle  d^2 \rangle }{3 e^2 {\lambda }_c } k + ...
\end{equation}
can be derived immediately from eq.s (3.6) and (3.8). Moreover from eq.s (3.8) 
and (3.6) it follows that, for $ k \rightarrow \infty $, $ \hat{f} (k) $ 
vanishes as $ k^{-3} $.  \par

Finally, using the result of eq.(3.8) in eq.(3.4), we obtain for the energy 
$ {\cal{E} }_0 $ the result 
%3.10
\begin{eqnarray}
&& {\cal{E} }_0 [ \hat{f} ] = E_0 - {\pi }^{-1} e^2 {\lambda }_c
 \int_0^{\infty } dk \, k \hat{f} (k) {\hat{n} }_e (k)   \\ \nonumber
&& = E_0 - {\pi }^{-1} e^2 {\lambda }_c \int_0^{\infty } dk \, k
 {\hat{n} }_e^2 (k)  ( 1 + \frac{2 \langle  d^2 \rangle }{3 e^2 {\lambda }_c } 
 k )^{-1}  ,
\end{eqnarray}
with $ \hat{f} $ given by eq.(3.8). Eq.(3.10) shows that $ {\cal{E} }_0 
[\hat{f} ] <  E_0 $, as expected.  \par 

We observe that the second (negative) contribution in eq.(3.10) would have the effect of shifting every energy-level of the atom by the same amount. Therefore such an effect cannot be observed through any spectral-line analysis.  \par

As an example, for the hydrogen atom in the 1s state we have 
%3.11
\begin{equation}
{\phi }_0 (r) = (\pi a^3 )^{-1} e^{-r/a}   ,
\end{equation}
an correspondingly
%3.12 
\begin{equation}
e^{-2} \langle  d^2 \rangle = 3 a^2   ,
\end{equation}
where $ a = {\hbar }^2 / m e^2 $ is the Bohr radius. Putting $ z_k =
\frac{1}{2}  ka $, we have
%3.13
\begin{equation}
{\hat{n} }_e (k) = (1 + z_k^2 )^{-2} \sim 1 - \frac{1}{2} k^2 a^2 +...
\end{equation}
In this case from eq.(3.8) we obtain
%3.14
\begin{eqnarray}
&& \hat{f} (k) = (1+ z_k^2 )^{-2} (1 + 4 {\alpha }^{-1} z_k )^{-1}  \\ \nonumber
&& = 4{\alpha }^{-1} (1 + z_k^2 )^{-2} (z_k + \alpha /4 )^{-1}  ,
\end{eqnarray}
where $ \alpha = e^2 / \hbar c $ is the fine-structure constant.  \par

Let us evaluate how many photons are present in the state (3.1). From 
eq.(2.1) one obtains
%3.15
\begin{equation}
U a_{\bm{k} \alpha } U^{-1} = a_{\bm{k} \alpha } + i
 \sqrt{\frac{2 \pi }{\hbar c k V}} e^{-i \bm{k} \cdot \bm{R} } \hat{f} (k) 
 \bm{d} \cdot {\bm{e} }^*_{\bm{k} \alpha }
\end{equation}
and a similar expression for $ a^+_{\bm{k} \alpha } $. In this way we obtain
%3.16
\begin{eqnarray}
&& n^{ph}_{\bm{k} } = \sum_{\alpha } \langle  g|\langle  0| 
U a^+_{\bm{k} \alpha } a_{\bm{k} \alpha } U^{-1} |0 \rangle|g \rangle   
  \\ \nonumber
&& = \frac{2 \pi }{\hbar ckV} {\hat{f} }^2 (k) [ \langle  d^2 \rangle - k^{-2}
 \langle  ( \bm{d} \cdot \bm{k} )^2 \rangle ]  . 
\end{eqnarray}

In the continuus limit, by integration over the direction of the wave-vector 
$ \bm{k} $, one obtains
%3.17
\begin{equation}
n_{ph} (k) = \frac{V}{8 \pi } \int d {\Omega }_k \, k^2 n^{ph}_{\bm{k} }
 = \frac{\langle  d^2 \rangle}{3 \pi \hbar c} k {\hat{f} }^2 (k)  ,
\end{equation}
with a total number of photons given by
%3.18
\begin{equation}
N_{ph} = \int_0^{\infty } dk \, n_{ph} (k) 
 = \frac{\langle  d^2 \rangle}{3 \pi \hbar c} \int_0^{\infty } dk \, k 
 {\hat{f} }^2 (k)  .
\end{equation}

For a hydrogen atom eq.(4.12) gives the result 
%3.19
\begin{eqnarray}
&& N_{ph} = \frac{{\alpha }^3 }{4 \pi } \int_0^{\infty } dz \, 
 \frac{z}{ (1 + z^2 ) (z + \alpha / 4 )^2 }   \\ \nonumber
&& \simeq 1.4 \times 10^{-7} .
\end{eqnarray}
For the calculation of the integral see the Appendix A.  \par

It is interesting to note that the maximum value for $ n_{ph} $ is attained 
for $ z =z_0 = 1/7 $, {\it e.g.} for $ \lambda \sim 22 a $, in the 
far-ultraviolet region. \par

The transformation $ U $ of eq.(2.1) could be introduced in the free-electron case also, in order to seek for a possible spontaneous localisation for the 
wave-packet, due to the second term in eq.(3.10). For {\it e.g.} a Gaussian 
$ {\phi }_0 \sim e^{-r^2 / a^2 } $, eq.(3.10) would give
 $ \Delta E_0 \sim - \frac{e^2 }{a} (\frac{{\lambda }_c }{a} )^2 $. But for a 
non-relativistic electron this negative contribution to the total energy would 
be too small to have any observable effect.

\section{N atoms}
In the case of $N$ identical atoms, the unitary transformation $U$ of eq.(2.1)
becomes
%4.1
\begin{equation}
U = \exp [ - \frac{i}{\hbar c} \sum_{a=1}^N {\bm{d} }_a \cdot {\bm{A} }^f 
 ({\bm{R} }_a ) ]   .
\end{equation}
As a consequence, eq.(2.6) now reads
%4.2
\begin{eqnarray}
&& [U E_{\bot i} U^{-1} ] (\bm{r} ) = E_{\bot i} (\bm{r} )  \\ \nonumber
&& - \frac{4 \pi }{V} \sum_a \sum_j \sum_{\bm{k} } e^{i \bm{k} \cdot 
 (\bm{r} - {\bm{R} }_a )} \hat{f} (k) ({\delta }_{ij} - \frac{k_i k_j }{ k^2 } )
 d_{aj}   .
\end{eqnarray}
The Hamiltonian $H_M $ of eq.(1.3) is replaced by
%4.3
\begin{eqnarray}
&& H_M = \sum_a [\frac{1}{2m} : \vert -i \hbar {\bm{\nabla }}_a + \frac{e}{c}
 \bm{A} ({\bm{r} }_a ) \vert^2 :   \\ \nonumber
&& + V( \bm{r} - {\bm{R} }_a ) ] + \sum_{a < b} \sum_{ij} d_{ai} \:
 g_{ij} ({\bm{R} }^{ab} ) \: d_{bj}   .
\end{eqnarray}  
In eq.(4.3) we have put $ {\bm{R} }^{ab} = {\bm{R}}_a - {\bm{R}}_b $. 
Furthermore the matrix $g$ is given by the expression
%4.4
\begin{eqnarray} 
&& g_{ij} (\bm{R} ) = R^{-3} ({\delta }_{ij} - 3 \frac{X_i X_j }{R^2 } )
       \\ \nonumber
&& \simeq - \frac{4 \pi }{V} \sum_{\bm{k} } e^{i \bm{k} \cdot \bm{R} }
 ({\delta }_{ij} - \frac{k_i k_j }{k^2 } )   .     
\end{eqnarray}
The last expression is correct for $ R > 0 $.  \par

In eq.(4.3) the last term represents an interaction, due to the Coulomb field
(longitudinal electric field), among the ED moments belonging to different 
atoms. In the standard treatment, this term alone is responsible for the van der
Waals interaction. In second-order perturbation theory it gives rise to an
interaction-potential, whose leading contribution at large distance is
proportional to $ R^{-6} $. However we will show that the unitary transformation
of eq.(5.1) gives rise to a similar interaction term, to be ascribed to the
transverse field. In fact we have
%4.5
\begin{eqnarray}
&& U H_T U^{-1} = H_F \\ \nonumber
&&   +  \sum_a \{ \frac{1}{2m} : \vert -i \hbar
 {\bm{\nabla }}_a + \frac{e}{c} [ \bm{A} ({\bm{r} }_a ) - {\bm{A} }^f 
 ({\bm{R} }_a ) ] \vert^2 :   \\ \nonumber
&&  + V ({\bm{r} }_a - {\bm{R} }_a )
 -{\bm{d} }_a \cdot {\bm{E} }^f_{\bot } ({\bm{R} }_a )
 + \frac{2}{3 \pi } | {\bm{d} }_a |^2 \int_0^{\infty } dk \,
 k^2 \hat{f} (k)    \\  \nonumber
&& - 4 \pi e^2 {\lambda }_c \sum_{\bm{k} } k^{-1} \hat{f} (k)
 e^{i \bm{k} \cdot ({\bm{r} }_a - {\bm{R} }_a )}  \\ \nonumber
&& + {\pi }^{-1} e^2 {\lambda }_c  \int_0^{\infty } dk \, k {\hat{f} }^2 (k) \}
      \\ \nonumber
&& + \sum_{a < b} \sum_{ij}  d_{ai} {\Gamma }_{ij} ({\bm{R} }^{ab} ) d_{bj}   .
\end{eqnarray}
The last term represents the interaction among the ED moments referring to
different atoms, due both to the Coulomb field and to the {\it modified}
transverse electric-field $ {\bm{E} }^f_{\bot } $. The matrix $ \Gamma $ is 
given by
%4.6
\begin{eqnarray}
&& {\Gamma }_{ij} (\bm{R} )    \\ \nonumber
&& = - \frac{4 \pi }{V} \sum_{\bm{k} } e^{i \bm{k}
 \cdot \bm{R} } [1 - {\hat{f} }^2 (k)] ({\delta }_{ij} - \frac{k_i k_j }{k^2 } )
       \\ \nonumber
&& = ({\delta }_{ij} {\nabla }^2  - \partial_i  \partial_j )
 \frac{4 \pi }{V} \sum_{\bm{k} }  k^{-2} [1 - {\hat{f} }^2 (k)] 
  e^{i \bm{k} \cdot \bm{R} }  \\ \nonumber
&& \simeq  ({\delta }_{ij} {\nabla }^2  - \partial_i  
 \partial_j ) \frac{2}{\pi } \int_0^{\infty } dk \,
 [1 - {\hat{f} }^2 (k) ] \frac{\sin kR}{kR}  \\ \nonumber
&& = ({\delta }_{ij} {\nabla }^2  - \partial_i  
 \partial_j ) \frac{2}{\pi R} \int_0^{\infty } dt \,
 [1 - {\hat{f} }^2 (\frac{t}{R} ) ] \frac{\sin t}{t}   .
\end{eqnarray}

The leading contribution to this expression can be evaluated by use, in the
calculation of the last integral, of the asymptotic expansion for $ \hat{f} $,
given in eq.(3.9). We obtain
%4.7 
\begin{eqnarray}
&& \frac{2}{\pi R} \int_0^{\infty } dt \,[1 - {\hat{f} }^2 (\frac{t}{R} ) ]
 \frac{\sin t}{t}     \\ \nonumber
&& \simeq \frac{8 \langle  d^2 \rangle }{3 \pi e^2 {\lambda }_c } R^{-2}  
\int_0^{\infty } dt \, \sin t = \frac{8 \langle  d^2 \rangle }{3 \pi e^2
{\lambda }_c } R^{-2}  ,
\end{eqnarray}
where the two contributions proportional to $ R^{-1} $ cancel each other. In 
this way, for the leading contribution to $ {\Gamma }_{ij} $ we obtain the
expression
%4.8
\begin{equation}
{\Gamma }_{ij} (R)  \sim \frac{16 \langle  d^2 \rangle }{3 \pi e^2 
{\lambda }_c } R^{-4} (4 \frac{X_i X_j }{R^2 } - {\delta }_{ij} )  .
\end{equation}
For example in the case of a pair of hydrogen atoms eq.(4.8) reads 
%4.9
\begin{equation}
{\Gamma }_{ij} (R) \sim \frac{16}{ \pi \alpha a^3 } (\frac{a}{R} )^4
 (4 \frac{X_i X_j }{R^2 } - {\delta }_{ij} )  .
\end{equation}

The interaction due to the last term in eq.(4.5) is responsible for the van der
Waals interaction. This will be analysed in the next section.  \par

Quite relevant in the $N$-atom case is the fact that the photon number of 
eq.(3.18) is multiplied by $N$. In fact a uniform distribution of $N$ atoms in a
volume $V$ can be assumed as a rough representation for a refractive medium. In
such a case, the number of photons per unit-volume would be given by
%4.10
\begin{equation}
\frac{N_{ph} }{V} \simeq  \frac{N}{V} \frac{\langle  d^2 \rangle}{3 \pi \hbar
c}  \int_0^{\infty } dk \, k {\hat{f} }^2 (k)   ,
\end{equation} 
with the following distribution with respect to the frequency $\omega $
%4.11
\begin{equation}
\frac{n_{ph} (\omega )}{V} \simeq \frac{N}{V} \frac{\langle  d^2 \rangle}{3 \pi
\hbar c^3 } \; \omega  {\hat{f} }^2 (\frac{\omega }{c} )    .
\end{equation}

Assuming $ N/V \sim 10^{23} \; {cm }^{-3} $ and using the result of eq.(3.19) 
one obtains $ N_{ph} / V \sim 10^{16} \; {cm }^{-3} $. This result (very 
roughly) represents the total number of photons per cubed centimeter, which are 
present actually in a refractive medium in its ground state.

\section{van der Waals potential}

The long-distance interaction between two neutral atom is due  essentially to 
the van der Waals force. Up to the second order the perturbation theory gives an
explicit expression for the dipole-dipole interaction energy.The standard
calculation is based on the Coulomb interaction as given by the last term of 
eq.(4.3), with $ g_{ij} $ given in eq.(4.4). The well-known result \cite{5} is
%5.1
\begin{eqnarray}
&& U' (R)   \\ \nonumber
&& \simeq \frac{6}{R^6 } \sum_{b_1 b_2 } \frac{|\langle  b_1|d^{(1)}_z 
|g \rangle |^2 \; |\langle  b_2|d^{(2)}_z |g \rangle |^2 }{2 E_0 
- E_{b_1 } -E_{b_2 } }   ,
\end{eqnarray}
which is proportional to $ R^{-6} $.   \par 

However in Sec.IV we have shown that the complete dipole-dipole interaction, is
represented by the last term in eq.(4.5), where the matrix $ g_{ij} $ is 
replaced by $ {\Gamma }_{ij} $ of eq.(4.8) (or by eq.(4.9) for a pair of 
hydrogen atoms). As a consequence, the potential $ U' $ of eq.(5.1) is replaced 
by
%5.2
\begin{eqnarray}
&& U (R) \simeq  \frac{11}{R^8 } \; \frac{16 \langle  d^2 \rangle}{3 \pi e^2 
{\lambda }_c }     \\  \nonumber
&& \sum_{b_1 b_2 } \frac{|\langle  b_1|d^{(1)}_z |g \rangle |^2 \;
 |\langle  b_2|d^{(2)}_z |g \rangle |^2 }{2 E_0 - E_{b_1 } -E_{b_2 } }   ,
\end{eqnarray}
which is proportional to $ R^{-8} $.   \par 

The ratio of the two expressions is
%5.3
\begin{equation}
U (R) / U' (R) \simeq \frac{11}{6 R^2 } \; [\frac{16 \langle d^2 \rangle}{3 \pi 
e^2 {\lambda }_c } ]^2    .
\end{equation} 

For two hydrogen atoms eq.(5.3) reads
%5.4
\begin{equation}
\frac{U (R)}{U' (R)} \simeq \frac{11}{6} (\frac{a}{R})^2
 (\frac{16}{\pi \alpha })^2\simeq 8.9 \times 10^5 (\frac{a}{R})^2   .
\end{equation} 

Let us recall that both $ U $ and $ U' $ are negative. From eq.(5.4) one can see
that $ |U(R)| > |U'(R)| $, for $ R < 9.4 \times 10^2 a $. Actually the
experimental status does not look conclusive.

\section{Conclusion}

We have shown that, in the ED representation, the correct Hamiltonian describing
an atom interacting with the radiation field is given by eq.(2.24), with  $
\hat{f} $ given by eq.(3.8). Such an expression appears sensibly more
complicated than the simple expression $ -\bm{d} \cdot {\bm{E} }_{\bot } $, 
which is assumed usually.   \par

We have analysed some consequences concerning namely  the photon-distribution 
and the van der Waals interaction. In the last case we have obtained the 
remarkable result that, at large distance, the van der Waals potential between
two atoms is proportional to $ R^{-8} $. We are not aware of any direct measurement of the van der Waals potential, between two neutral polarizable 
atoms, expecially at large distance. A wide bibliography on the van der Waals interaction can be found in Ref. \cite{6}.

\appendix

\section{Integrals}

We outline the method used to calculate the integrals referring to the hydrogen
atom.   \par

The integrals are either of the following type
%A.1
\begin{equation}
I_n (A) = \int_0^{\infty } dz \, (z^2 + A )^{-n}     ,
\end{equation} 
or of the following one
%A.2
\begin{equation}
I_{nm} (A,B) = \int_0^{\infty } dz \, (z^2 + A )^{-n} (z+B)^{-m}   ,
\end{equation}
with $ n,m $ positive integers.    \par

For $ n > 1 $ we have
%A.3
\begin{equation}
I_n (A) = (-)^{n-1} [(n-1)!]^{-1} \frac{{\partial }^{n-1} }{\partial A^{n-1} }
 I_1 (A)   ,
\end{equation}    
with
%A.4
\begin{equation}
I_1 (A) = \int_0^{\infty } dz \, (z^2 + A )^{-1} = \frac{\pi }{2 \sqrt{A} }  .
\end{equation}

In a similar way, for $ n \geq 1 $ and $ m \geq 1 $ we have
%A.5
\begin{eqnarray}
&& I_{nm} (A,B)       \\ \nonumber
&& = \frac{(-)^{n+m-2} }{(n-1)!(m-1)!}
 \frac{{\partial }^{n-1} } {\partial A^{n-1} }
 \frac{{\partial }^{m-1} } {\partial B^{m-1} } I_{11} (A,B)  .
\end{eqnarray} 
The integral $ I_{11} $ can be calculated by standard techniques. The result is
%A.6 
\begin{eqnarray}
&& I_{11} (A,B) = \int_0^{\infty } dz \, (z^2 + A)^{-1} (z + B)^{-1} 
   \\ \nonumber
&& = (A + B^2 )^{-1} [\frac{\pi B}{2 \sqrt{A} } + \frac{1}{2} \ln A - \ln B ].
\end{eqnarray} 

At the end of the calculations one has to put $ A=1 $ and $ B= \alpha /4 $.

% \begin{acknowledgments}

% \end{acknowledgments}

\end{document}